\documentclass[aps,prl,twocolumn,showpacs]{revtex4}
\usepackage[dvips]{graphicx}

\begin{document}

\title{Sharp transition for single polarons in the one-dimensional
Su-Schrieffer-Heeger model}
\author{D. J. J. Marchand$^1$,  G. De Filippis$^2$, V. Cataudella$^2$,
M. Berciu$^1$, N.~Nagaosa$^{3,5}$, N. V. Prokof'ev$^{4,6}$, A. S.
Mishchenko$^{5,6}$, and P. C. E. Stamp$^1$ } \affiliation{ $^1$
Department of Physics and Astronomy, University of British Columbia,
     Vancouver, BC, Canada, V6T~1Z1 \\
$^2$ CNR-SPIN and Dip. di Scienze Fisiche - Universit\`{a}
     di Napoli Federico II - I-80126 Napoli, Italy\\
$^3$Department of Applied Physics, The University of Tokyo, 7-3-1
     Hongo, Bunkyo-ku, Tokyo 113, Japan \\
$^4$Department of Physics, University of Massachusetts, Amherst,
     Massachusetts 01003, USA \\
$^5$Cross-Correlated Materials Research Group (CMRG), ASI,
      RIKEN, Wako 351-0198, Japan \\
$^6$RRC ``Kurchatov Institute'' - 123182 - Moscow - Russia
}

\begin{abstract}
We study a single polaron in the Su-Schrieffer-Heeger (SSH) model
using four different techniques (three numerical and one
analytical). Polarons show a smooth crossover from weak to strong
coupling, as a function of the electron-phonon coupling strength
$\lambda$, in all models where this coupling depends only on phonon
momentum $q$. In the SSH model the coupling also depends on the
electron momentum $k$; we find it has a sharp transition, at a
critical coupling strength $\lambda_c$, between states with zero and
nonzero momentum of the ground state. All other properties of the
polaron are also singular at $\lambda = \lambda_c$. This result is
representative of all polarons with coupling depending on $k$ and
$q$, and will have important experimental consequences (eg., in
ARPES and conductivity experiments).
\end{abstract}

\maketitle

Polarons have been of broad interest in physics ever since they were
introduced in 1933, to describe dielectric charge
carriers~\cite{landau33}. Apart from their central role in
solid-state physics, with many models now in use
\cite{frohlich,holstein,biopol}, they exemplify in quantum field
theory the passage from weak to strong coupling in a non-trivial
model of a single particle coupled to a bosonic field \cite{QFTpol}.
The first serious non-perturbative studies by Feynman \cite{RPFpol}
of the Frohlich polaron, are now a classic, but only recently were
accurate results established across the whole range of coupling
strengths \cite{Froh}. Since then, exact numerical studies have been
made of, eg., D-dimensional Holstein polarons in various lattice
geometries, with $D = 1,2,3$ \cite{KornTrug}; of 3D Rashba-Pekar
polarons with short-range interactions \cite{RP02}; of pseudo
Jahn-Teller polarons \cite{MN01}; and so on.

A central question in this field has been whether a sharp transition
can exist in the polaronic ground state as a function of the
dimensionless effective particle-boson coupling $\lambda$. In all
the above-cited work there is simply a smooth crossover, expected
when the coupling depends only on the bosonic momentum $q$; then
there must always be non-zero matrix elements between the ground
state and excited polaron states \cite{GL}. However, quite
generally, one expects the coupling to depend on both $q$ and the
particle momentum $k$; and then much less is known.

In this paper we study a specific example of this general case.  The
particle-boson coupling is taken from the well-known "SSH model",
introduced to describe electrons in 1-d polyacetylene \cite{SSH}.
Here we focus on the single polaron limit, not the more common case
of half-filling, and the bosons are chosen to describe optical
phonons. While this ignores the acoustic phonons which exist in real
materials, it allows a direct comparison with the large number of
results known for models which have a purely $q$-dependent coupling.
The Hamiltonian thus takes the simple form ${\cal H} = H_o + V +
H_{ph}$, where
\begin{equation}
H_o = -t_o \sum_i ( c_{i}^{\dagger} c_{i+1} + h.c.) \;\; \equiv\;\;
\sum_k \epsilon_k c_k^{\dagger} c_k \;,
 \label{Ho}
\end{equation}
describes the hopping of electrons between sites, with band
dispersion $\epsilon_k = -2t_o \cos(k)$ ($c_{i}^{\dagger}$ creates
an electron on site $i$; $c_k^{\dagger}$ creates
a momentum state $k$). The term $H_{ph} = \omega_{ph} \sum_i
b_i^{\dagger} b_i$ describes dispersionless phonons ($b_i^{\dagger}$
creates a phonon on site $i$). The interaction is
\begin{eqnarray}
V &=& -\tilde{\alpha}t_o  \sum_{i} ( \hat{X}_{i} -\hat{X}_{i+1})
(c_{i}^{\dagger} c_{i+1} + h.c.)
\nonumber\\
&=& N^{-1/2} \sum_{k,q} M(k,q) c_{k+q}^{\dagger} c_k (b_{-q}^{\dagger}+
b_q)
 \label{int}
\end{eqnarray}
with site displacements $\hat{X}_i = \sqrt{\hbar\over 2M\omega_{ph}}
\left(b_i+b_i^\dagger\right)$, and an interaction vertex
\begin{eqnarray}
M(k,q) &=& 2i \alpha [\sin(k+q)-\sin(k)] \nonumber \\
 &=& i(2\lambda \omega_{ph}t_o)^{1/2} [\sin(k+q)-\sin(k)]
 \label{vertex}
\end{eqnarray}
This interaction, with associated energy ${\alpha} =
\tilde{\alpha}t_o\sqrt{\hbar\over 2M\omega_{ph}}$, describes the
modulation of the hopping amplitude by phonons. We henceforth set
$t_o =1$, and define two dimensionless parameters: the
electron-phonon coupling parameter $\lambda= 2 \alpha^2/(t_o
\omega_{ph})= \langle |M(k,q)|^2\rangle/(2t_o \omega_{ph})$, where
$\langle \cdot\rangle$ averages over the Brillouin zone; and the
'adiabaticity' ratio $\omega_{ph}/t_o$ ($\equiv \omega_{ph}$ when
$t_o = 1$).

\vspace{2mm}

{\bf (i) Results}: We treat this non-perturbative problem with the
Momentum Average (MA) analytical approximation
\cite{Berc06,Good08,Berc10} and three different numerical
techniques: the Diagrammatic Monte Carlo (DMC) \cite{Froh}, the
Limited Phonon Basis Exact Diagonalization (LPBED) \cite{Fil09}, and
the Bold Diagrammatic Monte Carlo (BDMC)
\cite{ProkofevSvistunoz2008} methods. Applications of the first three
methods to polaron problems are well documented. However, our
implementation of the BDMC method for the SSH model contains several
new elements, reviewed in the supporting material.

\begin{figure}
        \includegraphics[scale=0.99]{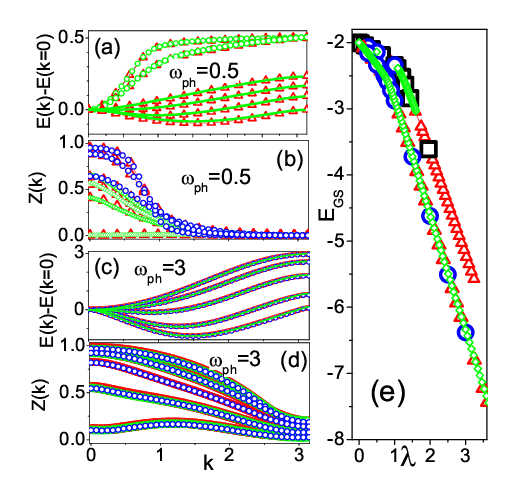}
\caption{(Color online) The polaron dispersion relation $E(k) -
E(k=0)$ is shown in (a) and (c), and the GS $Z$-factors $Z(k)$ at
momentum $k$ are shown in (b) and (d). Red (blue) triangles
(circles) correspond to LPBED (BDMC) methods. In (a), (b), where
$\omega_{ph}=0.5$, $\lambda = 0.25, 0.5, 1.0, 1.094, 1.21, 1.96$
(from top to bottom. In (c), (d), where $\omega_{ph}=3$, $\lambda =
0.25, 0.5, 1.0, 2.0, 4.0$ (from top to bottom). MA results are shown
as green solid curves. In (e) the GS energy for $\omega_{ph}=0.5$
(upper line) and $\omega_{ph}=3$ (lower curve) is shown; triangles,
rhombi, squares and circles correspond to LPBED, MA, DMC, and BDMC
methods, respectively. } \label{fig1}
\end{figure}

In the following we display results as functions of $\lambda$ in
both the adiabatic regime (choosing $\omega_{ph}=0.5$), and the
non-adiabatic regime (choosing $\omega_{ph}=3.0$). We begin with the
quasiparticle dispersion $E(k)$ and renormalization factor $Z(k)$
(Figs 1(a)-(d)). One sees immediately that whatever the
adiabaticity, the minimum of $E(k)$ is at $k=0$ for small $\lambda$,
but at finite $k$ for large $\lambda$. At first glance,
nevertheless, nothing unusual seems to happen to the ground state
energy $E_{GS}(\lambda)$ at the critical value $\lambda_c$, where
$k_{GS}$ first becomes non-zero (Fig. 1(e)). In fact, the curves in
Fig. 1(e) look quite similar to those for Holstein polarons.

\begin{figure}
        \includegraphics[scale=0.77]{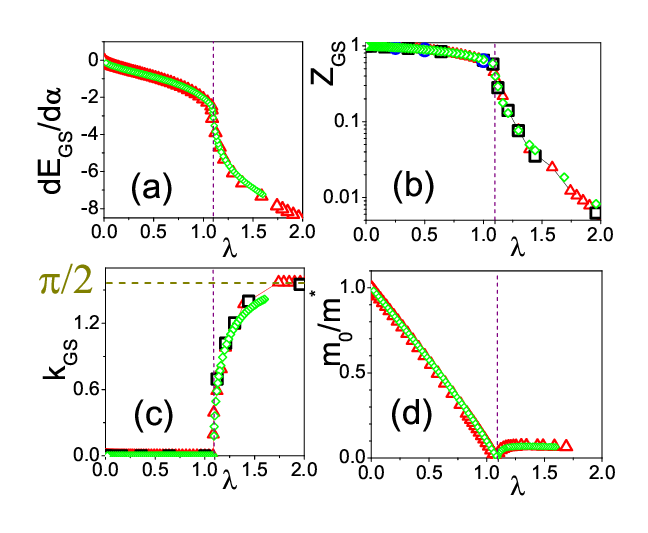}
\caption{(Color online) (a) Derivative of the GS energy with respect
to $\alpha$, (b) $Z$-factor of the GS, (c) wave vector of the GS,
and (d) the ratio $m_o/m^*$ of the bare and effective polaronic
masses at $k_{GS}$ for $\omega_{ph}=0.5$ (here $m_o = 1/2t_o$). Red
triangles, green rhombi, black squares and blue circles correspond
to LPBED, MA, DMC, and BDMC methods, respectively. The vertical
dashed arrow indicates the critical coupling $\lambda_c$. }
\label{fig2}
\end{figure}

However, there is actually a singularity at $\lambda_c$. Plots of
the dimensionless derivative $dE_{GS}(\lambda)/d\alpha$ (Fig. 2(a)),
the overlap $Z_{GS}(\lambda)$ between the ground state at finite
$\lambda$ and the uncoupled ground state (Fig. 2(b)), the momentum
$k_{GS}(\lambda)$ for which $E(k)$ is minimized (Fig. 2(c)), and the
renormalized effective polaron mass $m^*(\lambda) = (\partial^2
E(\lambda)/\partial k^2)^{-1}|_{k = k_{GS}}$ (Fig. 2(d)), all show a
sharp transition at $\lambda = \lambda_c(\omega_{ph})$ (see Fig. 2
for $\omega_{ph} = 0.5$, Fig. 3 for $\omega_{ph} = 3$). At this
singularity, the polaronic mass $m^*(\lambda)$ diverges, with
corresponding jumps in the first derivatives $d k_{GS}(\lambda)/ d
\lambda$  and $d Z_{GS}(\lambda) / d \lambda$, and in $d^2
E_{GS}(\lambda) / d \lambda^2$. However the average number of
phonons $N_{ph}(\lambda)$ in the polaronic polarization cloud does
not diverge at $\lambda_c$ (although it is presumably still
singular); $N_{ph}(\lambda_c) < 15$ for all values of the
adiabaticity parameter $\omega_{ph}$ checked so far. Note also how
$\lambda_c$ varies with $\omega_{ph}$ (Fig. 4), initially increasing
for small $\omega_{ph}$, but then falling to the asymptotic value
$\lambda_c \to 1/2$ in the instantaneous phonon limit $\omega_{ph}
\to \infty$. This limit can be derived analytically (see below).

\begin{figure}
        \includegraphics[scale=0.77]{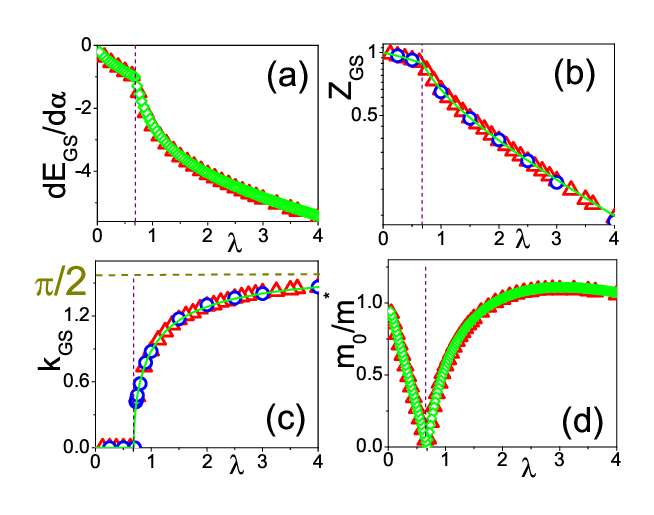}
\caption{(Color online) The same as Fig.~\ref{fig2} but for
$\omega_{ph}=3$. MA results in (b), (c) are shown as green solid
lines. } \label{fig3}
\end{figure}

We emphasize here the remarkable agreement obtained between all 4
methods. The three numerical techniques are in principle exact, but
all have their practical limitations, such as the sign problem noted
below for QMC methods.

\begin{figure}
        \includegraphics[scale=0.6]{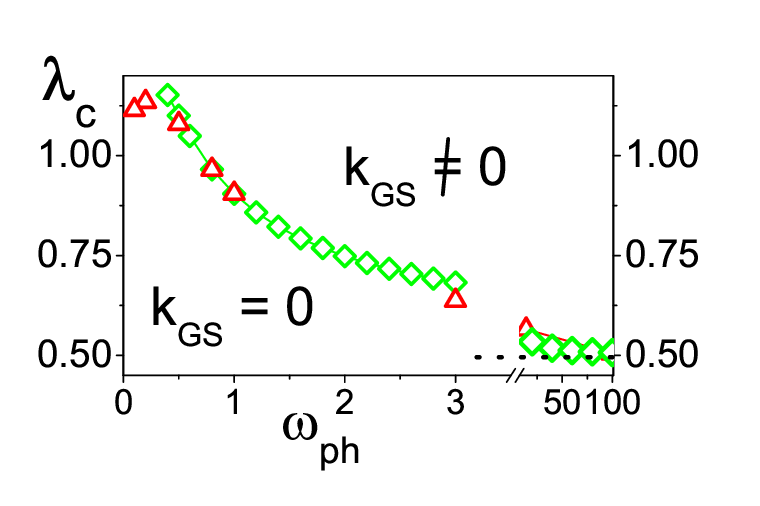}
\caption{(Color online) Phase boundary dividing GSs with zero and
nonzero momentum. Green squares and red triangles refer to MA and
LPBED methods. The horizontal dotted line refers to the
instantaneous limit $\lambda_c^{\omega_{ph} \to \infty} = 1/2$. }
\label{fig4}
\end{figure}

\vspace{2mm}

{\bf (ii) Discussion:} The key new feature of couplings $M(k,q)$
like that in Eqs. (\ref{int}) and (\ref{vertex}), compared to
$k$-independent couplings, is that they are non-diagonal in site
index. Thus phonons cause the bandwidth to fluctuate, and can by
themselves generate hopping between sites. The lowest-order process
contributing to $E_{GS}$ is 2nd order in $M(k,q)$; higher
corrections come from even powers of $M$. The same applies to the
polaron self-energy, the polaron mass, quasiparticle
renormalization, etc. Consider now a pair of vertices, connected by
a phonon of momentum $q$; we have
\begin{eqnarray}
 \label{pair}
M_{k,\, -q}M_{k' - q,\, q} & \propto & \lambda
\sin^2{\Big(\frac{q}{2}\Big)} \nonumber \\
& &  \times \cos{\Big(k - \frac{q}{2}\Big)}\cos{\Big(k' -
\frac{q}{2}\Big)}.
\end{eqnarray}

Three key new features appear in (\ref{pair}):

(a) it can be of either sign when $k \neq k'$. This leads to a 'sign
problem' in any Monte Carlo calculation (indeed, for any interaction
$M(k,q)$ with non-definite sign); we discuss this in the supporting material.
 The SSH model is thus representative of a
large class of models in which non-diagonal couplings give a sign
problem.

(b) multi-site hopping terms involving phonons generate terms in the
polaron dispersion of form $E(k) = E_0 - 2 t_1^*\cos k -
2t_2^*\cos(2k)-...$. Now for $q$-only dependent couplings, the
nearest-neighbor hopping $t_1^*\ll t_o$ is exponentially suppressed,
and $t_2^* \sim {t_o^2\over \omega_{ph}}e^{-{4\lambda t \over
\omega_{ph}}}$~\cite{Mars95} is doubly suppressed, because each
requires an intersite polaron cloud overlap. Here, however, an
electron can hop from $i-1 \rightarrow i \rightarrow i+1$, using
only $V$, to first create and then remove a phonon at site $i$. The
associated energy is $t_2^* \propto \Delta t_{i,i+1} \Delta
t_{i,i-1}/\omega_{ph}$ where $ \Delta t_{i,i-1}\sim \alpha$ is the
phonon-induced change in the hopping. Since the phonon-induced
displacement $\hat{X}_i$ increases one bond length while decreasing
the other, $\Delta t_{i,i+1} \Delta t_{i,i-1}<0$, ie., $t_2^*\propto
- \alpha^2/\omega_{ph}$ is {\it negative}, favoring a minimum in
$E(k)\sim - 2 t_2^* \cos(2k)$ at $k=\pi/2$, consistent with our
results for large $\lambda$. This simple analysis indicates how the
transition can occur. Of course, higher order terms must also be
considered; and a transition like this, signalled by the change in
$k_{GS}$, is certainly not guaranteed for all $k$-$q$-dependent
couplings (thus the Edwards model in the large $\lambda$ limit also
has a dominant $t_2^*$ term of similar origin; but $t_2^*>0$, and
$k_{GS}=0$ for all $\lambda$ \cite{Berc10}).

(c) Finally, consider the limit $\omega_{ph} \rightarrow \infty$ for
a fixed $\lambda, t_o$. The phonon propagator tends to its static
limit: $D(q,\omega)=-2\omega_{ph} / (\omega_{ph}^2-\omega^2) \to
\tilde{D} = -2/\omega_{ph}$.  The polaron propagator is then
dominated by the 2nd-order correction in $V$, scaling like
$\alpha^2/ \omega_{ph}\sim \lambda t_o$ (higher order corrections
$\sim \alpha^{2n}/\omega_{ph}^{n-1} \sim t_o\lambda^{n-1}
\left(t_o\over \omega_{ph}\right)^{n-1} \rightarrow 0$. Thus, to
lowest order in $\omega_{ph}^{-1}$ we get from Eq. (\ref{pair}) that
\begin{eqnarray}
E(k) &=& -2t_o \cos k + {1 \over 2N^{1/2}} \tilde{D} \sum_q \left|
M(k,q) \right|^2 \nonumber \\
&=& -2t_o \cos k - \frac{2 \lambda t_o}{\sqrt{N}} \sum_q [\sin(k+q)-\sin
k]^2 \;\;
 \label{inst}
\end{eqnarray}
 We see that the dispersion curvature
$\left[ d^2 E(k) / d k^2 \right]_{k=0} = 4t_o (1/2 - \lambda)$ at
$k=0$. Thus,  in the large $\omega_{ph}$ limit, the effective mass
diverges for $\lambda_c (\omega_{ph} \to \infty) = 1/2$. Fig. 4
shows it converges very slowly to this limit.

This discussion shows, at least for sufficiently large
$\omega_{ph}$, that there must be a critical coupling strength
$\lambda_c$ at which $k_{GS}$ leaves zero. For small $\omega_{ph}$
the existence of a critical point is less clear, because the higher
order diagrams can have arbitrary sign; but it is what we find here
for all $\omega_{ph}$ studied \cite{sun09}. Note, however, that for
$\omega_{ph}<0.3$, the average number of phonons $N_{ph}$ increases
significantly, making numerical simulations very difficult. The MA
method is also questionable in this limit.

One is tempted to call this $T=0$ transition a 'quantum phase
transition'. However this is not correct, because any phase
transition must involve the cooperative behaviour of an infinite set
of degrees of freedom; but here the number of phonons $N_{ph}$ in
the polaronic cloud always remains small. Of course with a
macroscopic number of polarons in the system, we would see
non-analyticity in bulk properties like $dE_{GS}(\lambda)/d\lambda$;
but a small number of polarons will be invisible in any
thermodynamic property. Thus we simply assert the existence of a
non-analyticity, as a function of $\lambda$, in the polaronic
properties.

We see that polarons having a coupling to a bosonic field depending
on both $k$ and $q$ behave in a fundamentally different way from the
standard case with only $q$-dependent coupling. This suggests a
large zoology of so far unexplored behavior in many physically
relevant systems. Note how surprisingly different the polaronic
properties are here. For example, for large $\lambda$,
$m^*(\lambda)$ {\it decreases}, and $Z(k_{GS})$ remains quite large.
We see that "standard polaronic behavior" is really just a feature
of models like the Holstein and Frohlich model.

Experimental signatures of the new behavior - notably, the critical
point - will clearly be invisible in any thermodynamic measurements.
However the divergence of the effective mass should be easily
detectable in transport measurements; the polaron mobility $\mu \sim
1/m^*$ goes to zero at the critical point. Thus in any system where
the charge mobility is carried by the polarons, this critical point
should be very obvious. It would also be interesting to do ARPES
experiments \cite{damas03}, where polarons can be ejected directly
from the insulating state, allowing direct measurement of $E(k)$ and
$Z(k)$. Apart from polyacetylene, various organic semiconductors are
known to have important non-diagonal coupling to phonons
\cite{orgSemi}, as do several dimerized Mott magnetic semiconducting
oxides \cite{mott}; in some of these, the coupling can be varied
somewhat by pressure. However, any quantitative theory for such
experiments must also include the coupling to longitudinal phonons,
electron-electron interactions, and inter-chain coupling.

{\bf (iii) Acknowledgements:} We thank G.A. Sawatzky for
discussions. NVP acknowledges support from NSF grant (PHY-1005543). NN
acknowledges a MEXT Grant-in-Aid for Scientific Research
(19048008, 19048015, 21244053). ASM is partly supported by RFBR (10-02-00047a).
VC and GDF acknowledge a University of Napoli "Federico II" FARO grant
(CUP-E61J10000000003). DM, MB and PCES acknowledge NSERC, CIFAR, FQRNT and PITP for
support.


{\center{\large{\bf SUPPORTING MATERIAL}}}

\vspace{3mm}

{\bf (i) Quantum Monte Carlo Methods:} As discussed in the main
text, a key feature of the problem addressed here is that any
Quantum Monte Carlo (QMC) method is faced by a sign problem (whereas
in the conventional polaron problem, with $q$-dependent coupling
only, the imaginary time polaron Green's function and self-energy
are sign-definite). Since the sign change can only happen when the
internal particle line changes momentum, we see that non-crossing
diagrams are positive-definite and only a fraction of the crossing
diagrams will have an overall negative sign. This is of course not
as severe as the sign-alternating series found in the many-body
fermionic case.

The Bold Diagrammatic Monte Carlo (BDMC) technique \cite{ProkS08} is a
sign-problem tolerant method for many-body problems. It is thus a
natural fit to the SSH polaron, and is one of the four
techniques used in this letter. It improves on the simpler Green's
Function Diagrammatic Monte Carlo\cite{ProkS98} (G-DMC)and
Self-Energy Monte Carlo ($\Sigma$-DMC) methods. The former is a
Monte Carlo sampling of the conventional Feynman diagram expansion
for the Green's function $G(k,\,\tau)$ where $k$ and $\tau$ are the
momentum and imaginary time respectively. The $\Sigma$-DMC method
also expands the self-energy $\Sigma(k,\,\tau)$ diagrammatically,
but contains fewer terms, thus providing faster convergence. The
BDMC method further reduces the number of diagrams to be sampled by
self-consistently renormalizing the particle propagator, and using
this to draw subsequent diagrams. By repeated iteration of this
procedure, the number of diagrams accounted for by the BDMC method,
grows exponentially with simulation time, instead of linearly as in
other QMC methods.

\begin{widetext}

\begin{figure}[hbt]
\includegraphics[scale=0.7]{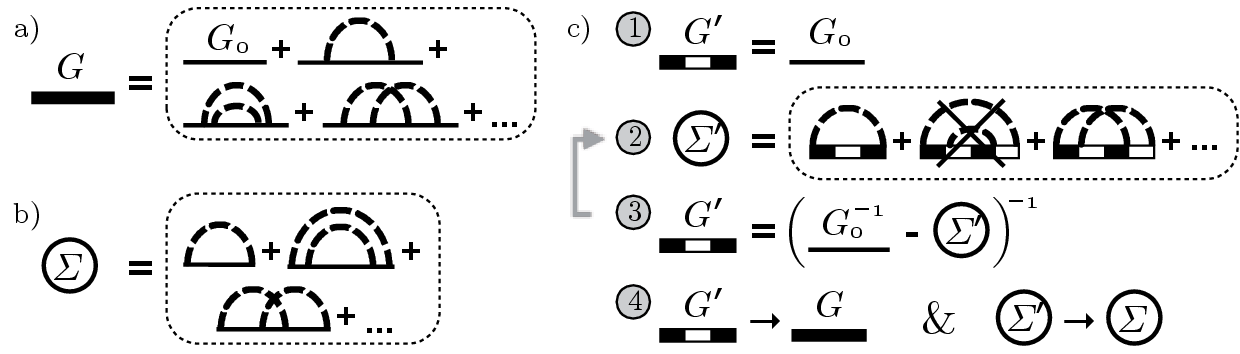}
\caption{\label{fig_DMC}Comparison between the (a) G-DMC, (b)
$\Sigma$-DMC and (c) BDMC methods. Dashed-lined boxes represent MC
sampling of the appropriate diagram expansion. Some extra
restrictions apply in the case of BDMC to avoid double counting.
Step 2 and 3 of BDMC are repeated until desired convergence. Step 3
consists in solving Dyson's equation in imaginary frequency.}
\end{figure}

\end{widetext}

In both $\Sigma$-DMC and BDMC the renormalized propagator $G'(k,\,\tau)$
is obtained by solving Dyson's equation in imaginary frequency $\xi$
using a FFT algorithm to convert between $\tau$- and $\xi$-dependent
quantities. Normalization of $\Sigma(k,\,\tau)$ is enforced using
${\lim\atop{\tau\to0}} \Sigma(k,\,\tau) =\lambda\omega t
[2-\cos{(2k)}]$. Quantities such as the polaron dispersion,
quasiparticle weight and effective mass can all be obtained either
from $\Sigma(k,\,\tau)$ or $G(k,\,\tau)$. Fig. \ref{fig_DMC}
compares the diagrammatic formulation of the various QMC methods.

\vspace{4mm}

{\bf (ii) The Momentum Average Method:} The Momentum Average (MA)
approximation solves the equation of motion for the Green's function
$G(k,\omega)= \langle 0 |c_k (\omega + i\eta - {\cal H})^{-1}
c_k^{\dagger} |0\rangle$, where $|0 \rangle$ is the vacuum. As usual, this
solution appears in the form of an infinite hierarchy of coupled
equations of motion. The main idea behind MA is to simplify the
coefficients multiplying the generalized Green's function in these
equations so that this infinite hierarchy can be solved
(quasi)analytically. Because it is formulated in terms of equations
of motion and because there is no truncation at any order, MA is a
non-perturbative approximation. In diagrammatic terms, MA is
equivalent \cite{Berc07} to a sum of all diagrams for the
self-energy expansion, after discarding exponentially small
contributions (compared to what is kept) from each diagram. As a
result, MA satisfies multiple spectral weight sum rules exactly.

The guidance on how to approximate the equations of motions (which
also determines what parts of the diagrams are discarded) comes from
the variational meaning of MA. Essentially, one makes some
assumption about the nature of the bosonic cloud, and keeps only
terms consistent with it \cite{Berc07,Bar07,Berciu07}. For the
Holstein model, it is already a very good approximation to assume
that all the bosons in the cloud are at the same site -- however,
there can be arbitrarily many such bosons, and the particle can be
arbitrarily far from the cloud. Further systematic increases in the
variational space lead to systematic improvements of the MA
approximation \cite{Berc07}. For more complicated models, one needs
to allow the cloud to be more extended. In this work, we use the
straightforward generalization to our "SSH model" of the 3-site
version of MA detailed for the Edwards model in Ref.
\cite{Berc10}.

{\bf (iii) The Limited Phonon Basis Exact Diagonalization method:}
The Limited Phonon Basis Exact Diagonalization (LPBED)
method\cite{defilippis} is based on the standard Lanczos algorithm.
It uses the translational symmetry associated with periodic boundary
conditions, requiring that the states have a definite momentum (the
Hamiltonian is block diagonalized). Each basis vector is a linear
superposition with appropriate phases of the translational copies
(charge carrier and lattice configurations are together rigidly
translated) of a state having the electron fixed at a site and
phonon quanta located around it. The real bottleneck comes from the
Hilbert space required by the phonon basis, which is unlimited even
in a finite size lattice. To circumvent this difficulty, LPBED takes
into account only a finite number of phonon states. The states are
chosen starting from the observation\cite{Bar07} that the MA
approximation within the Holstein model can be recovered by using a
restricted basis where all the bosons in the cloud are at the same
site and arbitrarily far from the electron. In the SSH model, where
the coupling charge-lattice is not local and is related to the
hopping, the phonon cloud has to be more extended to describe in an
appropriate way the relevant physical processes (the extension
increases as $\omega_{ph}$ decreases). In this work we use a cluster
of 5 neighbouring sites (the cluster can be arbitrarily far from the
electron). Moreover we include an additional pair of phonons located
on any pair of lattice sites. In this way the scattering processes
between the charge carrier and up to two phonons in $q$-space are
exactly treated, so that the weak coupling limit is optimally
described. The main advantage of the LPBED method is that it is
possible to calculate not only the self energy of the quasiparticle
but any correlation function.

\end{document}